\documentclass[linenumbers]{aastex631}

\newcommand{\nup}{$\nu_{\rm peak}^S$}

\usepackage{graphicx}

\begin{document}

\title{Changing Look of the optical spectrum of the MeV Blazar PKS~0446+112 (4FGL~J0449.1+1121)}

\author[0000-0002-2239-3373]{Simona Paiano}
\affiliation{INAF - IASF Palermo, via Ugo La Malfa, 153, I-90146, Palermo, Italy}

\author{Renato Falomo}
\affiliation{INAF - Osservatorio Astronomico di Padova, vicolo dell'Osservatorio 5, I-35122, Padova, Italy}

\author{Aldo Treves}
\affiliation{Universit\`a dell'Insubria, via Valeggio, 22100, Como, Italy}
\affiliation{INAF - Osservatorio Astronomico di Brera, via Bianchi 46, I-23807, Merate (Lecco), Italy}

\author{Riccardo Scarpa}
\affiliation{Instituto de Astrofisica de Canarias, C/O Via Lactea, s/n E38205, La Laguna (Tenerife), Spain }
\affiliation{Universidad de La Laguna, Dpto. Astrofsica, s/n E-38206, La Laguna (Tenerife), Spain }

\author[0000-0001-6620-8347]{Boris Sbarufatti}
\affiliation{INAF - Osservatorio Astronomico di Brera, via Bianchi 46, I-23807, Merate (Lecco), Italy}

\begin{abstract}
Following the high activity of the $\gamma$-ray \textit{Fermi} source 4FGL~J0449.1+1121 (PKS~0446+112), possibly associated with a IceCube neutrino event IC-240105A, we obtained optical spectroscopy with the Gran Telescopio Canarias of the counterpart. We detect a clear emission line at 3830 \AA\ identified as Ly$\alpha$ that confirms the redshift of source at z=2.153.
Comparing with previous spectroscopy, we find an increase of the continuum by a factor $\sim$10, and a significant decrease of the CIV 1550 emission line flux by a factor $\sim$5. This produces a dramatic drop of the Equivalent Width from $\sim$20\AA\ to 0.8\AA, which is suggestive of a very high jet activity. 
The Full Width Half Maximum of the emission lines are midway (1000 - 2000 km/s) between those typical of the broad and narrow regions of quasars. 
Based on this, the source classification is intermediate between Flat Spectrum Radio Quasar and BL Lac object.
\end{abstract}

\keywords{BL Lacertae objects: individual (PKS 0446+11) --- galaxies: distances and redshifts --- gamma-rays: galaxies --- neutrinos }


\section{Introduction} \label{sec:intro}

Blazars, radio-loud AGNs powered by supermassive black holes with relativistic jet axis pointing close to the direction of the observer, represent the vast majority of the $\gamma$-ray extragalactic sources. Their Spectral Energy Distribution (SED) is dominated by non-thermal radiation of the jet and is characterized by a doubled-humped shape with two broad peaks \citep[see e.g. ][ and references therein]{falomo2014}{}{}: the first peak located typically in the IR to X-ray region and attributed to synchrotron radiation, while for the second one, spanning in the hard X-ray to $\gamma$-ray band, the origin is still debated and explained as inverse Compton emission by electrons in the leptonic models, or in hadronic scenarios as due to synchrotron emission from protons or through photo-meson interactions.

PKS~0446+112 (a.k.a 4FGL~J0449.1+1121) is a prototype of “MeV Blazars” \citep[][and references therein]{ghisellini2009, sbarrato2015, marcotulli2017}. 
Their synchrotron peak is located in the far infrared, the Compton peak in the MeV range, and they are characterized by high $\gamma$-ray luminosities.

First observations of PKS~0446+112 were reviewed by \citet{halpern2003} where they identify as blazars a number of optical counterparts  of $\gamma$-ray detections by EGRET.
Their optical spectrum (4000~-~8000 \AA) is essentially featureless, with a marginally significant emission at 4880 \AA. 
No indication of a host galaxy was present in a CCD exposure, suggesting the source is at relatively high redshift.

The source enters in the \textit{Fermi}/LAT First Source Catalog \citep[1FGL, ][]{abdo2010a}, and in the first catalog of active nuclei detected by \textit{Fermi} \citep[1LAC, ][]{abdo2010b}. An optical spectrum of the source is reported by \citet{shaw2012} within a multi-year observing campaign to follow up \textit{Fermi} blazars \citep{healey2008}. 
The spectrum exhibits a clear emission line at $\sim$4880~\AA.
The identification with CIV~1550 was proposed yielding a redshift z~=~2.153. 

The spectral energy distribution (SED), using GROND, \textit{Swift}, \textit{NuSTAR} and \textit{Fermi}/LAT data, was studied by \citet{marcotulli2017}. The SED has the typical double humped shape of blazars, with the synchrotron peak in the infrared, indicating a low-energy synchrotron peak BL Lac object (LBL)\footnote{Blazars are categorized based on the rest-frame frequency of the low-energy synchrotron peak (\nup) into
low- (LBL: \nup~$<10^{14}$~Hz), intermediate- (IBL:
$10^{14}$~Hz$ ~<$ \nup~$< 10^{15}$~Hz) , and high-energy
(HBL: \nup~$> 10^{15}$~Hz ) peaked objects \citep[][]{padovani1995, abdo2010} }. The Compton peak is at $\sim$ 100 MeV. From the observation of the blue bump, following \citet{ghisellini2015} based on the standard \citet{shakura1973} accretion disk model, \citet{marcotulli2017} estimated a black hole (BH) mass $\sim$5.0$\times$10$^8$~M$_{\odot}$.

In November 2023, PKS~0446+112 was observed in a rather high $\gamma$-ray state \citep[][]{giroletti2023}, a factor $\sim$18 above the 4FGL-DR4 level \citep{4FGLDR4}.
A possible IceCube neutrino event associated with the source was also reported \citep[][]{gcn35485} on 2024, January 05.
This prompted an observational effort in all spectral bands, where high activity was detected \citep{atel16398, atel16399, atel16402, atel16407, atel16409, atel16417, atel16453}.

In this work, we report on two optical spectra taken with the Gran Telescopio Canarias (GTC) in February 2024 which are unprecedented for the covered spectral range, and for the signal to noise ratio.
This enables us to characterize and classify this peculiar source, as well as to appreciate substantial spectral changes.

\section{Observations and data analysis} \label{sec:obs}

PKS~0446+112 was observed with the Gran Telescopio Canarias (GTC) at the Roque de Los Muchachos on 2024 February, with the spectrograph OSIRIS \citep{cepa2003}. the spectrograph was configured with a slit width of 1.2" and two different grisms, R1000B and R2500U, covering the spectral range 3650~-~7780~\AA\ and 3445~-~4590~\AA\ and yielding spectral resolution R = $\lambda$/$\Delta$ $\lambda$ $\sim$600 and $\sim$1300 respectively. 

The strategy of the observations and the data reduction (carried out using IRAF software and standard procedures for long slit spectroscopy) are the same adopted in \citet{paiano2017tev}.
For each setting, we obtained three individual exposures, in order to remove cosmic rays and CCD cosmetic defects. 
All exposures were combined into a single average spectrum. 
The resulting spectra (with a signal-to-noise SNR$\sim$250 for the R1000B spectrum and SNR$\sim$25 for the R2500U one), exhibit both emissions and absorptions, with every single feature accurately reproduced across the individual spectra.

Wavelength calibration was performed using the spectra of Hg, Ar, Ne, and Xe lamps and the accuracy of the wavelength calibration is 0.1\AA. 
Spectra were corrected for atmospheric extinction using the mean La Palma site parameters.
Relative flux calibration was derived from the observations of a spectro-photometric standard star during the same observation night.
An absolute flux calibration of the spectra was done using the source magnitude  estimated by a direct $r$ filter image obtained as part of the target acquisition. Very significant variation in the observed magnitude (r=17.2) was found with respect to the SDSS value (r=19.9).
Finally, the spectra were de-reddened ( by applying the law of \citet{cardelli1989}, assuming the value of the Galactic extinction \textit{E(B–V)}=0.45 as from the NASA/IPAC Infrared Science Archive 6.7\footnote{https://irsa.ipac.caltech.edu/applications/DUST/}\citep{schlafly2011}.
The spectra are shown in Fig.\ref{fig:GTC_R1000} and Fig.\ref{fig:GTC_R2500}  and are 
also reported in ZBLLAC database (see \citet{landoni2020}).

\section{Results} \label{subsec:res}

\begin{figure*}[ht!]
    \centering
    \includegraphics[width=0.52\textwidth, angle=-90]{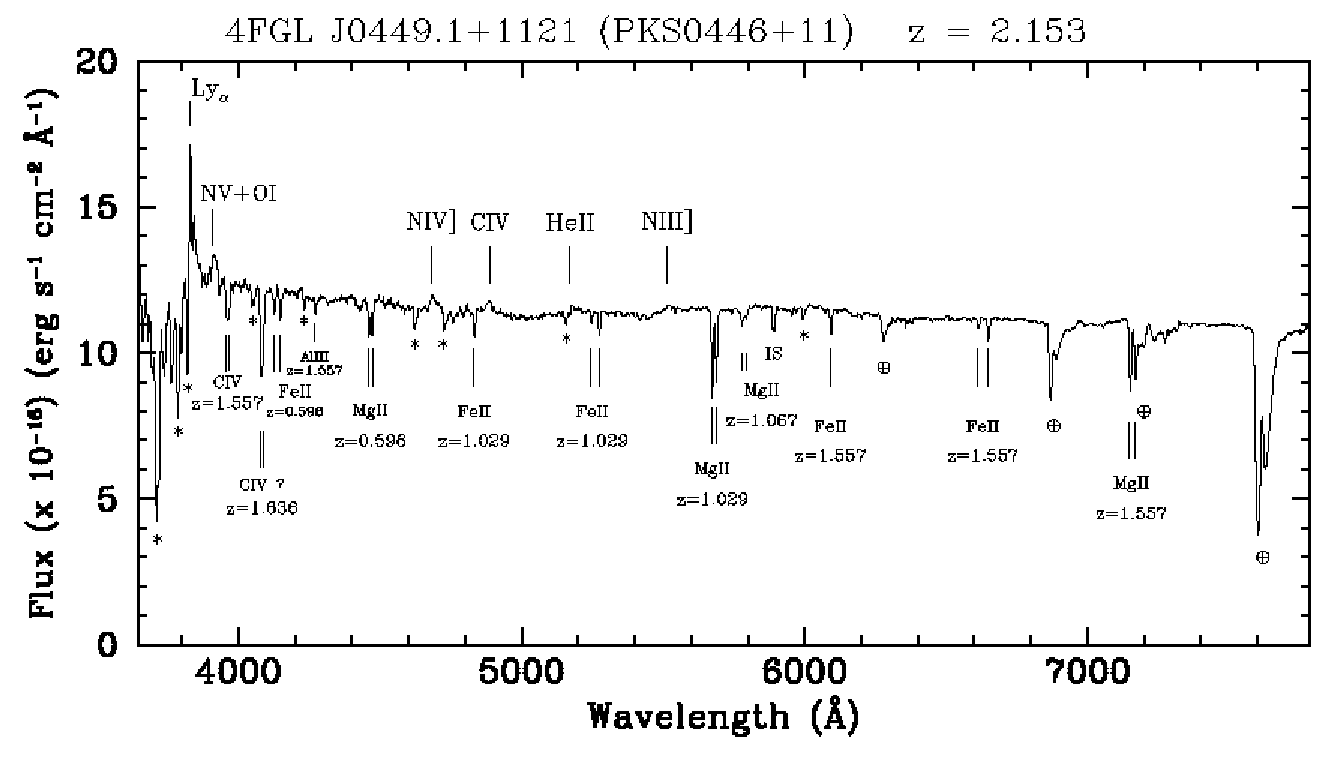}
  \caption{The optical spectrum of PKS~0446+112 obtained at GTC with OSIRIS+R1000B grism. The spectrum is corrected for reddening assuming E(B–V)=0.45. A prominent Ly$\alpha$ emission line is clearly detected at z~=~2.153. In addition other weaker emission lines are also present (see text and Tab. \ref{tab:opt_em_id}). The high quality spectrum shows a number of intervening absorption lines. Most of them belong to five different intervening systems (z~=~0.596, z~=~1.029, z~=~1.067, z~=~1.557, and z~=~1.636). The other unidentified intervening absorption lines are marked by *. Absorption feature due to interstellar medium is labeled as IS and the main telluric bands are marked by $\oplus$. }
\label{fig:GTC_R1000}
\end{figure*}

\begin{figure*}[ht!]
\centering
  \includegraphics[width=0.5\textwidth, angle=-90]{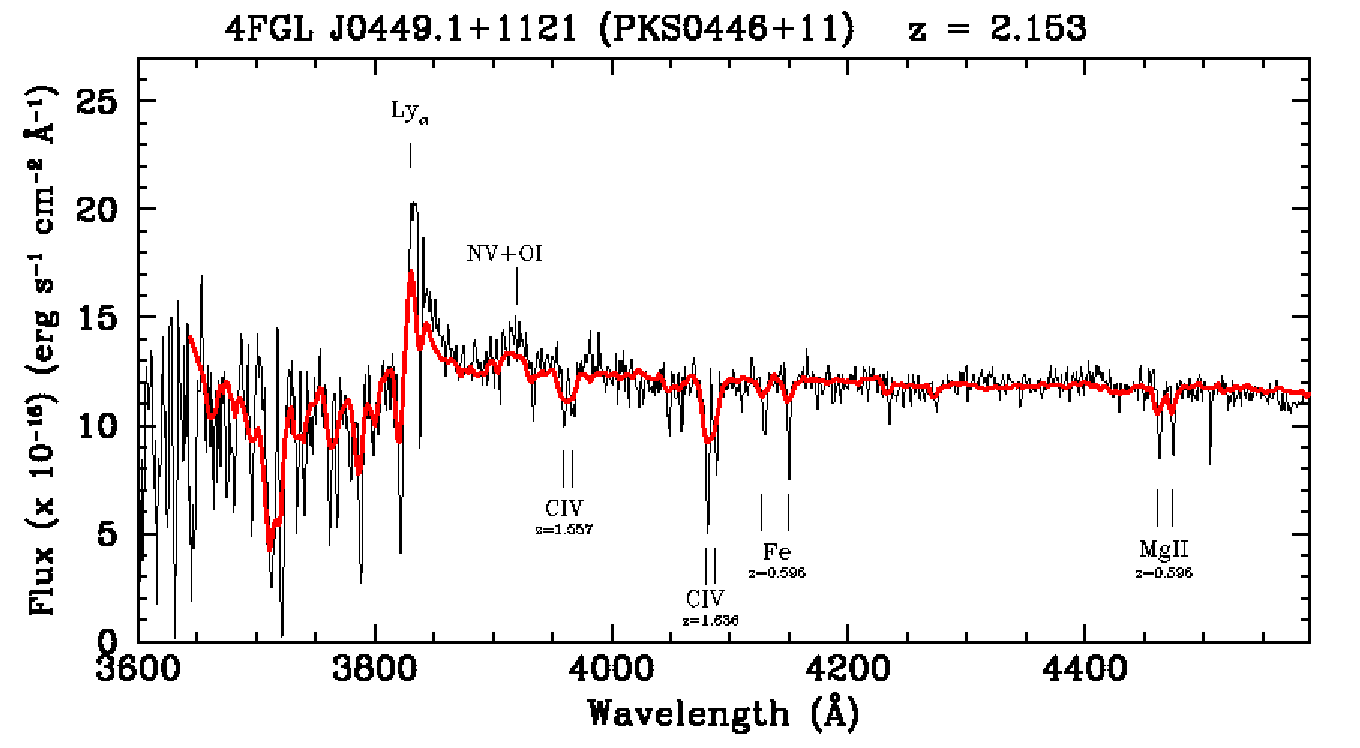}
    \caption{The R2500U optical spectrum of PKS~0446+112 obtained at GTC with OSIRIS, with the R1000B spectrum overlaid (red curve). The prominent Ly$\alpha$ emission line is detected and absorbed both in the blue and red parts.  The higher resolution spectrum allows us to resolve the other absorption features characteristic of the Ly$\alpha$ forest in the blue side and in the red part to resolve the doublet of the CIV intervening systems at z~=~1.557 and z~=~1.636. }
    \label{fig:GTC_R2500}
\end{figure*}

\begin{figure}
    \hspace{-1.6cm}
   \includegraphics[width=0.58\textwidth, angle=0]{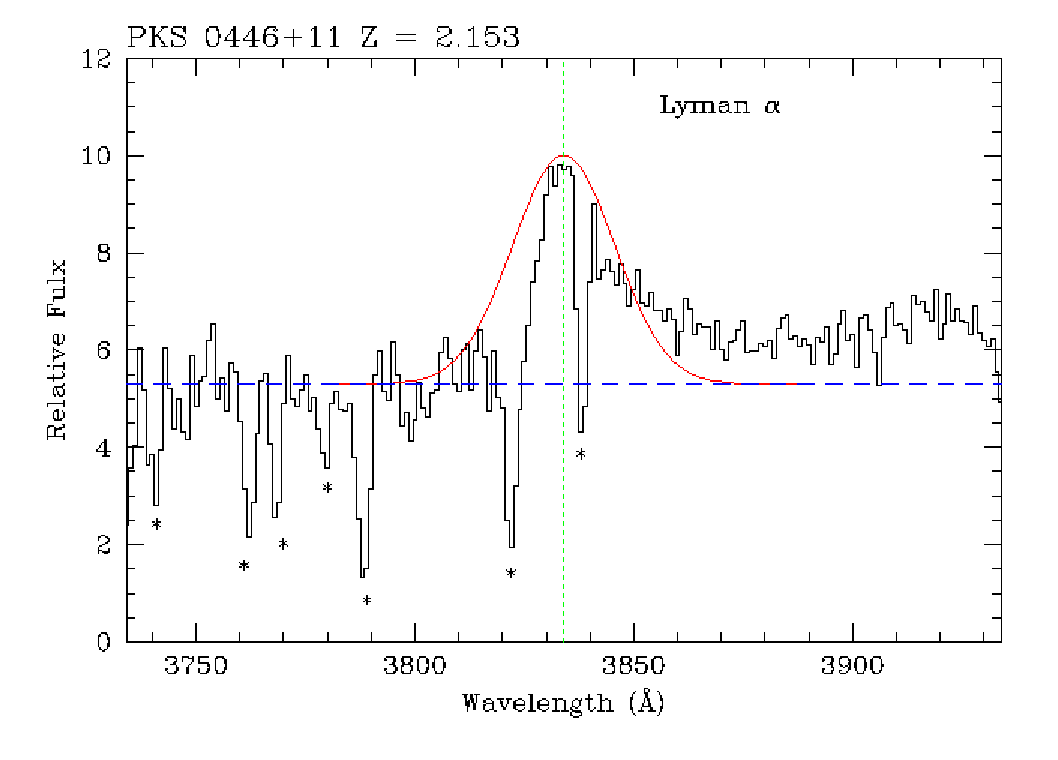}
    \caption{The optical spectrum obtained at GTC with OSIRIS+R2500U grism. The Ly $\alpha$ emission line 
 at z=2.153 is clearly detected. The red gaussian line represents the possible un-absorbed  emission line at equivalent width of 26 \AA\ and a FWHM of 2100 Km/s. The dashed blue line is the assumed continuum below the feature and the vertical dotted line gives the center. The main intervening absorption features are marked by *.}
    \label{fig:lyalfa}
\end{figure}

\begin{figure*}
    \centering
   \includegraphics[width=0.5\textwidth, angle=-90]{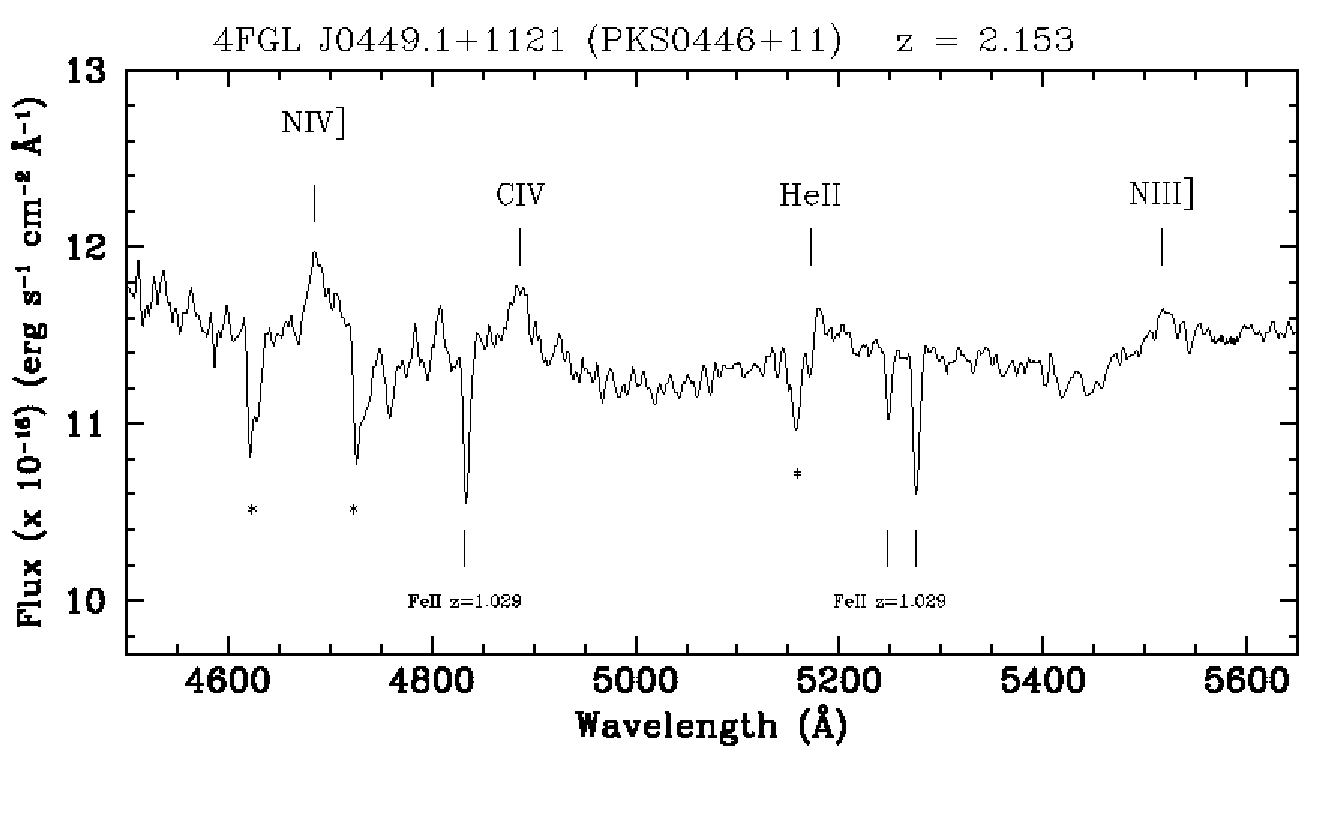} 
    \caption{Close-ups focused on the emission lines with EW$<$1 \AA\ detected in the medium resolution GTC optical spectrum (see Tab.\ref{tab:opt_em_id}). The unidentified intervening absorption features are marked by *. }
    \label{fig:GTC_emiss}
\end{figure*}

The medium resolution optical spectrum (S/N $\sim$250) obtained at GTC is reproduced in Figure \ref{fig:GTC_R1000}. The continuum  is characterized by a power law shape with spectral index $\alpha$ = -0.1 (F$_\lambda \sim \lambda^\alpha$). The most prominent feature visible in our spectrum is 
the emission line at $\sim$~3833~\AA\ that is identified as Ly$_\alpha$ 1215 . 
The line is absorbed in the blue (at $\sim$3822~\AA) and in the red (at $\sim$3839~\AA). Many other absorption features characteristic of Lyman forest are also visible on the blue side of Ly$_\alpha$ (see also Fig.\ref{fig:GTC_R2500}).
In Fig.\ref{fig:lyalfa} we show the Ly$_\alpha$ line together with a fit with a gaussian of Full Width Half Maximum (FWHM) = 2100 km/s, an un-absorbed Equivalent Width (EW) of 26~\AA\ and an integrated flux of F(Ly$\alpha$) $\sim$140$\times$10$^{-16}$ erg/cm$^2$ s$^{-1}$. 
This line confirms the redshift proposed by \citet{shaw2012} of z~=~2.153 that was based on the single emission line of CIV~1550.
In addition to Ly$_\alpha$ emission, we detect four other very weak (EW $<$ 1\AA) emission features at the same redshift  (see Tab.\ref{tab:opt_em_id} and Fig.\ref{fig:GTC_emiss}).\\

\begin{table}[ht!]
\begin{center}
\caption{Optical emission features of PKS0446+112 at $z$=2.153 } 
\begin{tabular}{cclccc}
\hline
$\lambda$ & EW$_{obs}$  &  Line ID  & Line Flux  ($\times$10$^{-16}$)    \\   
    (\AA) & (\AA)       &           & erg/(cm$^2$ s)  \\     
\hline
4685  & 0.9 $\pm$ 0.2   & NIV] 1486   & 10.3 $\pm$ 2\\ 
4887  & 0.8 $\pm$ 0.2   & CIV 1550    & 9.0  $\pm$ 2 \\ 
5173  & 0.30 $\pm$ 0.2  & HeII 1640   & 3.9  $\pm$ 2 \\ 
5518  & 0.25 $\pm$ 0.1  & NIII] 1750  & 2.9  $\pm$ 1 \\ 
\hline
\end{tabular}
\label{tab:opt_em_id}
\end{center}
\raggedright
\footnotesize{\textbf{Note.} Column 1: Observed central wavelength of the feature; Column 2: Observed equivalent width ; Column 3: Line identification, Column 4: Flux of the emission line. 
}
\end{table}

\begin{figure*}[ht!]
    \includegraphics[width=1.1\textwidth, angle=-90]{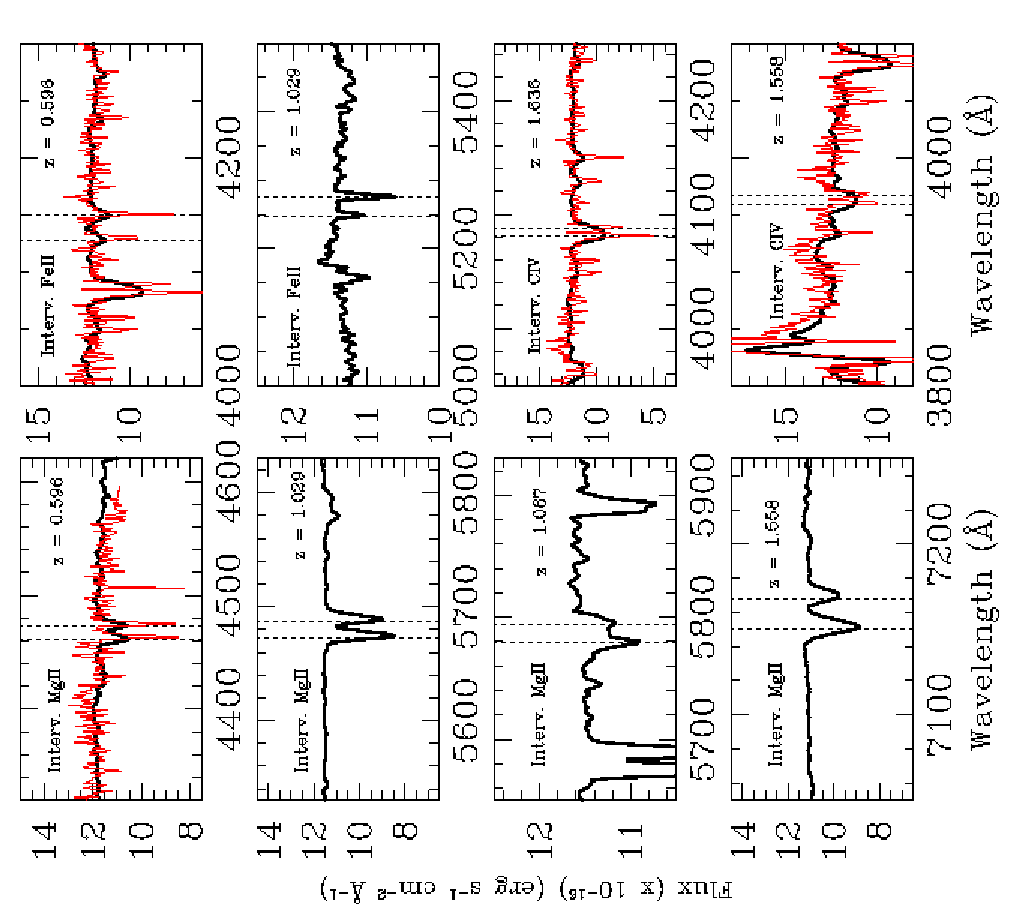}
    \caption{  Close-ups focused on the absorption doublet of MgII 2800 for four intervening absorption systems at z=0.596, z=1.029, z=1.067, and z=1.558 ({\it left panels}).  For all these systems, FeII absorption lines are also found (see in the {\it top right panels} ). The red spectrum refers to that obtained at higher spectral resolution (see text). 
    {\it Bottom right panels} The CIV 1550 intervening absorptions at z=1.636 and z=1.558.  }
    \label{fig:GTC_mg}
\end{figure*}

Besides the previously mentioned emission lines, our optical spectrum contains several weak absorption features due to intervening gas. There are at least five absorption systems at z=0.596, 1.0293, 1.067, 1.5578, and 1.6363 (see Fig.\ref{fig:GTC_R1000} and \ref{fig:GTC_mg}), and identified by the MgII 2796,2803, CIV 1548,1551, and FeII 2382 and FeII 2586,2600 lines (see details in Tab. \ref{tab:opt_abs_id}).
The most prominent absorptions are Mg II 2800 at $\sim$5680\AA\ with EW of $\sim$4 \AA\ and that of CIV 1550 at $\sim$4085 \AA\ with EW of $\sim$3\AA. The four MgII absorptions have rest-frame EW in the range 0.5 to 1.8\AA\ and are thus classified as strong system likely associated with massive galaxies or galaxy halos. Deep and high resolution near-IR images of the field could therefore detect the galaxies responsible of these intervening absorption features within few arcsec from the target.

\begin{table}
\begin{center}
\caption{Optical spectral absorption features associated to intervening systems.} 
\begin{tabular}{lccl}
\hline
$\lambda$  & EW   & Line ID  & $z$ \\  
(\AA) & (\AA)  &   &               \\   
\hline
4127      & 0.6 & FeII 2586   & 0.596  \\
4150      & 0.8 & FeII 2600   &   \\
4463      & 0.9 & MgII 2796  &   0.5963 \\ 
4474      & 0.7 & MgII 2803  &          \\  
\hline
4834      & 0.4 & FeII 2382  & 1.0293  \\ 
5248      & 0.2 & FeII 2586   &   \\ 
5276      & 0.5 & FeII 2600  &   \\ 
5674      & 2.2 & MgII 2796 &   \\ 
5688      & 1.7 & MgII 2803 &   \\ 
\hline
5779      & 0.7 & MgII 2796  & 1.067  \\ 
5794      & 0.4 & MgII 2803  &   \\ 
\hline
3959      & 0.8 & CIV 1548 & 1.5578    \\  
3967      & 1.0 & CIV 1551 &     \\ 
6094      & 0.6 & FeII 2382 &      \\
6615      & 0.3 & FeII 2586  &      \\ 
6650      & 0.6 & FeII 2600 &      \\ 
7152      & 1.7$^{*}$ & MgII 2796 &    \\
7170      & 0.9$^{*}$ & MgII 2803 &    \\
\hline
4081      & 1.9 & CIV 1548 &  1.6363  \\ 
4089      & 1.2 & CIV 1551 &   \\ 
\hline
\end{tabular}
\label{tab:opt_abs_id}
\end{center}
\footnotesize{\textbf{Note.} Column 1: Central wavelength of detected the absorption feature; Column 2: Observed equivalent width of the feature; Column 3: Feature identification, Column 4: Redshift of the intervening system.\\
$^{*}$ The absorption system is strongly contaminated by the telluric absorption band. \\
}
\end{table}

\section{Discussion} \label{subsec:conc}

\begin{figure*}
    \centering
    \includegraphics{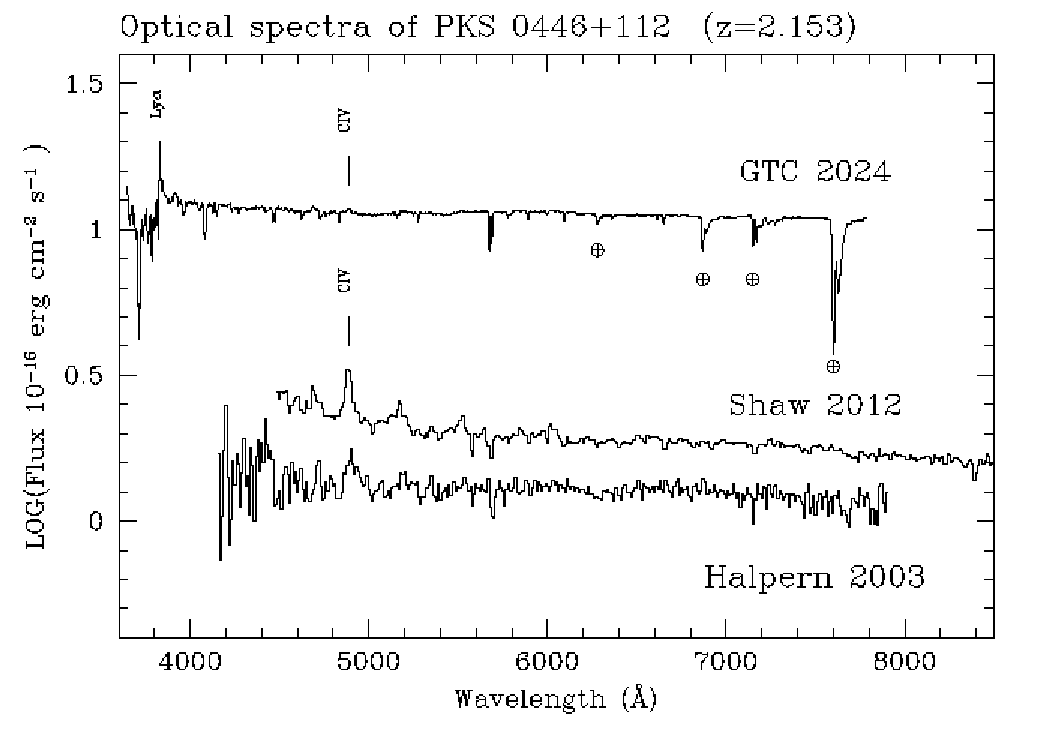}
    \caption{The optical spectra of PKS 0446+112 (4FGL J0449.1+1121) obtained at three epochs. Our spectrum obtained at GTC shows for the first time the Ly $\alpha$ emission together with many absorption features. Note the dramatic change of intensity of the CIV 1550 emission line (see also text). Main telluric bands are marked by $\oplus$. }
    \label{fig:allspec}
\end{figure*}

The optical spectrum of PKS~0446+112 exhibits a significant difference in the  features with respect to previous observations reported in literature. In particular there is a dramatic change of the intensity of the CIV 1550 emission line as displayed in the spectrum\footnote{The spectrum have been extracted and digitized from the Fig. 1 of \citet{shaw2012} } reported by \citet{shaw2012}.
We find a CIV 1550 line flux of 9$\times$10$^{-16}$ erg cm$^{-2}$ s$^{-1}$ while the spectrum obtained by \citet{shaw2012} yields 44$\times$10$^{-16}$ erg cm$^{-2}$ s$^{-1}$. This variability is illustrated in Fig.\ref{fig:allspec}. 
Another optical spectrum\footnote{The spectrum have been extracted, digitized and de-reddened, using the same prescriptions adopted for our GTC spectrum (see Sec. 2), from the Fig. 2 of \citet{halpern2003}} was obtained by \citet{halpern2003}. 
In this spectrum, with a lower SNR, the CIV emission line is barely visible. A rough estimate of its flux yields $\sim$36$\times$10$^{-16}$ erg cm$^{-2}$ s$^{-1}$, similar to that found by \citet{shaw2012} within the errors (see details in Tab.\ref{tab:civ}) 
There is therefore a  variation of a factor $\sim$5 in the intensity of the CIV 1550 line while the continuum varied by a factor 10.
The combination of a significant drop (by a factor $\sim$ 5) of CIV line intensity while there is an increase of the continuum by a factor 10, yields an impressive change of the EW by a factor $\sim$30. 
In the SDSS quasar sample considered by \citet{shen2011}, there is a statistically monotonous increase of the CIV line with the intensity of the continuum. The behavior of PKS~0446+112 is just opposite. It is similar to that of jet-dominated AGN B2 1420+32 \citep{mishra2021}, a changing look blazar at z=0.682. In this case, the MgII flux decreased by a factor $\sim$2, while the continuum increased by $\sim$10, yielding a variability\footnote{This is based on spectra with MgII measurements from \citep{mishra2021}} of the EW by a factor $\sim$10.

On the basis of the rest-frame EW of the emission lines ($\sim$8 \AA\ for the Ly$_\alpha$ and $\leq$0.3 for the other lines (see Tab.\ref{tab:opt_em_id}) ), PKS~0446+112 appears as an intermediate object between Flat Spectrum Radio Quasar (FSRQ) and BL Lac object (BLL), adopting the usual distinction between the two classes at 5 \AA. 
The modest values of the FWHM of the emission lines, $\sim$2100 km/s of Ly$_{\alpha}$ 1215 and $\sim$1200 km/s of CIV 1550, as well as for the other emission lines, are a factor $\sim$3 smaller than the average FWHM values of the emission lines in quasars reported by \citet{shen2011}. Moreover these values are closer to the value of 1200 km/s for separation between broad/narrow lines \citep{hao2005}. 
This may indicate a \textit{bona fide} BLL, rather than a masquerading BLL \citep{padovani2019}, namely objects with hidden broad emission lines swamped by a very bright, Doppler-boosted relativistic jet superimposed to the normal quasar spectrum.


 The substantial change of the EW of CIV 1550 emission line, and the opposite behaviour between the continuum and line fluxes, together with a variation of a factor $\sim$2 of the FWHM (see Tab.\ref{tab:civ}) indicates a rather complex interaction between the jet and the emission line regions. This is possibly due to some time delays and dynamic modifications of the emission region.

Finally we comment on the relationship between PKS~0446+112 and the detection of the IceCube neutrino event.
The association is not very strong ("bronze" event in the IceCube jargon), however, since the source was close to a phase of strong gamma and radio activity, the counterpart PKS~0446+112 seems favoured.
The neutrino event would be noticeable because of its high redshift. 
Since the source has the characteristic of LBL, it could be distinct from the class of blazars (IBL and HBL) which exhibits a statistically significant association ($\sim$4$\sigma$) with IceCube events \citep{giommi2020}.\\

\begin{table*}
\begin{center}
\caption{Measurements and properties from CIV emission line } 
\begin{tabular}{llcccc}
\hline
Spectrum & EW   & F$_{\lambda 4887}$  & FWHM   &  Line Flux & log(L$_{line}$) \\ 
        & (\AA) & (erg/(cm$^{2}$ s \AA)) & (km/s) & (erg/(cm$^{2}$ s)) & (erg/s)  \\ 
        &       &  $\times$10$^{-16}$      &        &  $\times$10$^{-16}$     &  \\   
\hline
\citet{halpern2003}   & 30 $\pm$ 5  & 1.2 & 3500 $\pm$ 600   & 36 $\pm$ 6 & 44.1  \\ 
\citet{shaw2012}      & 20 $\pm$ 3  & 2.2 & 2500 $\pm$ 400   & 44 $\pm$ 6.6 & 44.2  \\ 
This work (2024) & 0.8 $\pm$ 0.2 & 11.5 & 1200 $\pm$ 200   & 9 $\pm$ 2  & 43.5  \\ 
\hline
\end{tabular}
\label{tab:civ}
\end{center}
\raggedright
\footnotesize{\textbf{Note.} Column 1: Reference of the optical spectrum , Column 2: Observed Equivalenth Width of CIV emission line; Column 3: Flux of the continuum, Column 4: Full Width Half Maximum of the CIV emission line;  Column 5: Flux of the CIV emission line; Column 6: Luminosity of the CIV emission line.\\
}
\end{table*}

\section*{Acknowledgments}

We thank Paolo Padovani and Paolo Giommi for their valuable feedback and insightful comments on the discussion of the manuscript.



\bibliography{4FGLJ0449_R1_arxiv}{}
\bibliographystyle{aasjournal}



\end{document}